\documentclass[12pt]{iopart}
\RequirePackage{graphicx}
\usepackage{latexsym}
\expandafter\let\csname equation*\endcsname\relax
\usepackage{xcolor}
\expandafter\let\csname endequation*\endcsname\relax
\usepackage{amssymb}
\usepackage{amsmath} 
\usepackage{todonotes}
\begin{document}

\title{Fractional entropy of the Brown-Kucha\v{r} dust in fractional anti-de Sitter quantum gravity}

\author{ P. F. da Silva J\'{u}nior$^{1}$, S. Jalalzadeh$^{1,2}$\footnote{Author to whom any correspondence should be addressed}, and H. Moradpour$^3$}
\address{$^1$ Departamento de F\'{i}sica, Universidade Federal de Pernambuco, Recife-PE, 50670-901, Brazil }
\address{$^2$ 
Center for Theoretical Physics, Khazar University,
41, Mehseti Street, Baku AZ1096, Azerbaijan}
\address{$^3$ Research Institute for Astronomy and Astrophysics of Maragha (RIAAM), P.O. Box 55134-441, Maragheh, Iran}
 \ead{pedro.fsilva2@ufpe.br, shahram.jalalzadeh@ufpe.br,  h.moradpour@riaam.ac.ir}

\vspace{10pt}
\begin{indented}
\item[]March 2024
\end{indented}

\begin{abstract}
This study derives the mass spectrum and entropy of the Brown--Kucha\v{r} dust in anti-de Sitter (AdS) spacetime using the fractional Wheeler--DeWitt (WDW) equation. The generalized fractional WDW equation is formulated using a fractional quantization map, demonstrating a correlation between the fractal mass dimension of the Brown--Kucha\v{r} dust and Lévy's fractional parameter $\alpha$ of the Riesz fractional quantum operator. These findings may provide new insights into the ramifications of the fractal behavior of cosmic structures in quantum cosmology and quantum gravity.
\end{abstract}

%%\submitto{\CQG}
%\maketitle

\section{Introduction}\label{sec1man}

A quantum theory of gravity is one of the most challenging puzzles in theoretical physics, and nowadays several models are being investigated for a consistent resolution to the problem (for a review, see \cite{Jalalzadeh:2020bqu, Rovelli:2004tv, Oriti:2009zz}). On the one hand, quantum gravity may shed light on the singularities question
that appears in general relativity (GR), in both black holes \cite{Jalalzadeh:2022rxx} and the Big Bang context in the early universe \cite{Penrose:1964wq, Hawking:1970zqf}. Besides, quantum effects in gravity are expected to solve the black hole information paradox \cite{Hawking:1975vcx}. On the other hand, quantum gravity may give some important insights into the source of dark matter and dark energy \cite{Jalalzadeh:2023mzw, Jalalzadeh:2022dlj, Jalalzadeh:2022bgz, Trugenberger:2024miq, Chen:2024iha}.

The search for a complete theory was recently set aside in favor of explaining some phenomena at energy scales much lower than the Planck scale and offering potential quantum corrections to GR, nevertheless, a comprehensive theory and experimental evidence on the quantum aspects of gravity are still missing \cite{Kruglov:2023rii}. This kind of methodology aims to describe an effective quantum gravity (EQG) theory \cite{Moniz:2020emn, Brunetti:2022itx, Kelly:2020uwj}. In this regard, in both the settings of canonical quantum gravity \cite{Liu:2024soc} and quantum cosmology \cite{Bojowald:2012xy}, as well as in corrections to the level of gravitational action \cite{Donoghue:2012zc, Bertini:2024onw, Battista:2023iyu, Giacchini:2021abr}, various models of EQG have appeared over the last few years.

An emergent approach to EQG applies fractional calculus to quantum gravity in order to capture non-trivial quantum effects that lead to insights into fundamental problems in gravitation and cosmology \cite{LeonTorres:2023ehd,Shchigolev:2013jq,Calcagni:2013yqa,Calcagni:2017via,Shchigolev:2021lbm,Gonzalez:2023who,Socorro:2023ztq,Marroquin:2024ddg}. This methodology includes a variety of applications, such as a fractional interplay between dark matter and gravity \cite{Benetti:2024nxr}, holographic dark energy \cite{Trivedi:2024inb}, phenomenological quantum gravity \cite{Varao:2024eig}, implementation of a fractional action cosmology \cite{Jamil:2011uj,El-Nabulsi:2017vmp}, in discrete gravity \cite{El-Nabulsi:2013mma}, examining inflation and CMB spectrum \cite{Rasouli:2022bug}, in a chaotic inflation scenario \cite{El-Nabulsi:2013mwa}, using Ornstein--Uhlenbeck-like fractional differential equations in cosmology \cite{El-Nabulsi:2016dsj}, wormholes in fractional action cosmology \cite{El-Nabulsi:2017jss}, phantom cosmology with conformal coupling \cite{Rami:2015kha}, and even to analyze new metrics from fractional gravity \cite{El-Nabulsi:2017rdu}. Furthermore, fractional calculus is used to find the value of the cosmological constant more properly, due to the well-known ultraviolet divergence in ordinary quantum field theory \cite{Landim:2021www, Landim:2021ial}. Also, to explore features of the modified Newtonian dynamics (MOND) \cite{Varieschi:2020ioh, Giusti:2020rul}.

In particular, the authors of the fractional quantum gravity (FQG) model in references \cite{Moniz:2020emn, Rasouli:2021lgy} adopted Laskin's work \cite{Laskin:2002zz} on fractional quantum mechanics (FQM) to generalize the WDW equation of canonical quantization by a quantum Riesz fractional derivative operator. In FQM, Feynman's formulation of quantum mechanics was used by Laskin \cite{Laskin:2002zz} to generalize the path integrals by using Lévy's process instead of integrals built on Wiener's process. Lévy's processes are stochastic processes that generalize the description of Brownian motion, endowed with memories and a non-Gaussian distribution for the direction of the trajectory increments. As a consequence, the Brownian motion (Wiener process) is generalized in the sense that a parameter $\alpha$, called Lévy's fractional parameter, emerges to denote the fractal dimension of the trajectory, generally $d_{\text{fractal}}=\alpha$. As proposed by Laskin, with the intention of preserving the calculation of expected values of position and momentum in the fractional setup of quantum mechanics, $\alpha$ must be constrained to $1<\alpha{\;}{\leqslant}{\;}2$. 

 Besides that, fractional calculus, with its non-local nature and ability to capture long-range dependencies, such as big jumps and memory \cite{metzler2000}, provides a mathematical framework to represent and analyze fractal phenomena. In physical terms, it is not a new idea that the evolution of fractal systems has fractional derivatives, and fractional equations of motion, as a natural tool to be described \cite{rocco1999,butera2014}, or that fractional calculus can be used to change the fractal dimension of any random or deterministic fractal \cite{tatom1995}. In fact, the parameter $\alpha$, more generally ranging between $0<\alpha{\;}{\leqslant}{\;}2$, controls the heavy-tailed nature of the probability distribution Lévy flights and is very often associated with the fractal structure (and dimension) of physical systems on its own. Given the domain of quantum fluctuations, to cover quantum gravity effects one has no hope of observing well-defined results for expected values of operators, then $\alpha$ should vary as $0<\alpha{\;}{\leqslant}{\;}2$.

By using FQG \cite{Moniz:2020emn}, with the basic strategy of FQM \cite{Laskin:2002zz} and canonical quantization in a fractional quantization map \cite{Rasouli:2021lgy}, several developments for the potential resolution of fundamental questions in classical and quantum gravity, as well as in quantum cosmology (QC), have been achieved in recent years, as in black holes thermodynamics \cite{Jalalzadeh:2021gtq}, de Sitter cosmology \cite{Jalalzadeh:2022uhl}, inflation cosmology \cite{Rasouli:2022bug}, the emergence of cosmic space and dark matter problem \cite{Junior:2023fwb}, the synchronicity problem \cite{deOliveiraCosta:2023srx}, Friedmann equations in a fractal cosmological setting \cite{Jalalzadeh:2024qej}, and the dynamics of the stars from a fractional molecular cloud \cite{Moradpour:2024uqa}.  

Inspired by this approach, and taking into account the clue to a direct relationship between Lévy's fractional parameter and the fractal geometry associated with a fractal dimension of mass distribution, as indicated in ref. \cite{Moradpour:2024uqa}, this paper aims to investigate this relation between $\alpha$ and the fractal structure of a cosmological pressureless matter component (dust) provided by a scalar field filling a homogeneous and isotropic Friedmann--Lemaître--Robertson--Walker (FLRW) universe. The scalar field suitable for a description that comoving coordinates of the dust particles, as well as, the proper time along the dust worldlines, become canonical coordinates in the associated phase space is the Brown--Kucha\v{r} dust \cite{Brown:1994py}. With this prescription, canonical quantization in QC minisuperspace is implemented, and the reduced phase space is governed by a WDW equation in which the scalar field arising from the  Brown--Kucha\v{r} dust behaves as the time coordinate of a Schrödinger--WDW equation that can be solved by separation of variables. 

As a background to the canonical formalism of an FLRW universe, we have chosen to analyze the case of a model in the presence of a negative cosmological constant, $\Lambda<0$, i.e. an AdS spacetime. Although AdS spaces are unphysical in a direct cosmological sense (which is thought to resemble a de Sitter (dS) space due to the universe's observed accelerated expansion), it is a highly valuable model in theoretical physics for investigating elements of gauge theories \cite{Witten:1998zw, Dong:2016eik, DHoker:2002nbb} and quantum gravity \cite{Ryu:2006bv, Almheiri:2014cka, Faulkner:2013ica, Harlow:2018tng, Strominger:1998yg}. More importantly, AdS spacetime as a toy model becomes an essential tool for comprehending quantum features of spacetime because of the advent of the AdS/CFT correspondence \cite{Maldacena:1997re}, which builds a bridge between gravity in AdS spacetime and quantum field theories on its boundary. Specifically, the use of fractional operators for quantum cosmological models directly associated with the mass dimension of the associated mass distribution can affirm that the fractional quantum models of the considered EQG do in fact describe fractal effects in matter and its dynamics \cite{Moradpour:2024uqa}. The background of an AdS spacetime can then be useful for, among other applications, mapping this fractional-fractal method onto lower-dimensional spaces \cite{Basteiro:2022usu, Skvortsov:2017ldz}.

The content of this paper is organized as follows: In Section \ref{sec2man}, we derive the Hamiltonian formalism of the usual gravitational action for gravity minimally coupled with a Brown--Kucha\v{r} dust in an AdS spacetime. In section \ref{WDW}, the corresponding minisuperspace WDW equation is obtained in a Schrödinger-like equation for a quantum oscillator with the dust field representing the time coordinate, whose solution in the semiclassical limit for a flat ($k=0$) AdS spacetime gives us the mass spectrum and entropy of the dust. Finally, in Section 4, the fractional quantization map proposed is implemented to achieve the generalized fractional WDW equation in the presence of the Brown--Kucha\v{r} dust. The fractional mass spectrum and entropy are then examined in the context of the fractional parameter $\alpha$, and its relation with the fractal mass dimension is established. In our conclusion, in the last section, we present a final discussion of the results and prospects for future works.

\section{Quantum cosmology with Brown--Kucha\v{r} dust}\label{sec2man}

Let us consider a homogeneous and isotropic universe with a negative cosmological constant, and filled with a Brown--Kucha\v{r} dust. The metric for this perspective is given by the Robertson--walker line element
\begin{equation}\label{FR1}
\begin{split}
\mathrm{d}s^2&=-N^2(t)\mathrm{d}t^2+a^2(t)\Big\{\frac{\mathrm{d}r^2}{1-kr^2}+r^2(\mathrm{d}\theta^2+\sin^2(\theta)\mathrm{d}\varphi^2)\Big\}\\&=-N^2(t)\mathrm{d}t^2+a^2(t)h_{ij}\mathrm{d}x^i\mathrm{d}x^j,
\end{split}
\end{equation}
such that the lapse function, $N(t)$, is considered homogeneous, the shift vector is assigned a value of zero due to the isotropic space, $N^{i}=0$, $a(t)$ is the scale factor, and the curvature index, $k$, can be $k=1$, $k = 0$ or $k=-1$, for a closed, a flat, or an open spatial sections, $h_{ij}$, of the spacetime manifold $\mathcal M$, respectively. 

The usual Einstein--Hilbert action functional for gravity in the presence of the cosmological constant, and Brown--Kucha\v{r} dust is expressed by 
\begin{equation}\label{1-1aaa}
S=\frac{1}{16\pi G}\displaystyle\int_\mathcal{M}(R-2\Lambda)\sqrt{-g}\mathrm{d}^4x+S_\text{dust}+S_\text{GHY},
\end{equation}
where $S_\text{GHY}$ is the Gibbons--Hawking--York boundary term, $g$ is the determinant of the metric (\ref{FR1}), $\Lambda$ is the cosmological constant, and $S_\text{dust}$ is the action functional of the Brown--Kucha\v{r} dust as the considered matter field~\cite{Brown:1994py}. We assumed $\mathcal M$ to be a spatially compact, globally hyperbolic Lorentzian manifold.  Hence, the spacelike sections of the spacetime manifold possess a finite volume with the normalized curvature index $k$ \cite{doi:10.1142/8540}.
Different topologies are determined by different ranges of variation for the coordinates $(r, \theta, \varphi)$. 

Naturally, this is an extension of the internal interpretation in which time is not constructed from pure geometric data alone but from matter fields. To build time from matter, one can adopt a cloud of point-particle clocks covering all space, where each of these clocks measures time at points along its timelike worldline.  
For example, here we consider the Brown--Kucha\v{r} dust as this matter field~\cite{Brown:1994py}.  Such a matter field can be treated as a collection of non-canonical scalar fields $\rho$, $T$, $Z^a$, and $W_a~(a=1,2,3)$ which are equivalent to a dust (pressureless) fluid. The explicit form of action \ref{1-1aaa} for a minimally coupled Brown--Kucha\v{r} dust is
\begin{eqnarray}\label{dustabc}
S_\text{dust}=-\frac{1}{2}\int_{\mathcal M} \sqrt{-g}\rho(g^{\mu\nu}u_\mu u_\nu+1)\mathrm{d}^4x,
\end{eqnarray}
where $\rho$ is the mass density of the dust, and the velocity one-form, $u_\mu$, is defined by 
\begin{eqnarray}
u_\mu :=-\nabla_\mu T+W_a (\nabla_\mu Z^a).
\end{eqnarray}

The Euler--Lagrange equations corresponding to $T$, $Z^a$, and $W_a$ are given by
\begin{align}
\nabla_\mu (\rho u^\mu )=0,\qquad \rho  u^\mu (\nabla_\mu Z^a)=0,\qquad \nabla_\mu(\rho W_a u^\mu)=0, \label{eqU3}
\end{align}
respectively. In addition, the Euler--Lagrange equation corresponding to mass density, $\rho$, gives us the constraint on the norm of $u_\mu$ as $g^{\mu\nu}u_\mu u_\nu=-1$.
It is easy to check that the variation of metric, $\delta g^{\mu\nu}$, gives us the energy-momentum tensor of the dust fluid
\begin{align}
T^\text{(dust)}_{\mu\nu}=\rho u_\mu u_\nu. \label{em-bk}
\end{align}
Note that in obtaining the above equation, we have used $g^{\mu\nu}u_\mu u_\nu=-1$.

Now, in a homogeneous and isotropic FLRW spacetime, one can assume $Z^a=0$, $W_a= 0$, due to the isotropy of space and $\rho=\rho(t)$, $T=T(t)$, due to the homogeneous space. Therefore, the action (\ref{dustabc}) reduces to
\begin{equation}
    \label{dustaction}
    S_\text{dust}=-\frac{1}{2}\int_{\mathcal M}\sqrt{-g}(g^{\mu\nu}\partial_\mu T\partial_\nu T+1)\rho\mathrm{d}^4x=-\frac{\mathcal V_k}{2}\int Na^3\rho(t)\Bigg(-\frac{\dot T^2}{N^2}+1\Bigg)\mathrm{d}t,
\end{equation}
where $\mathcal V_k=\int_\Sigma\sqrt{h}\mathrm{d}^3x$ is the 3-volume volume of the spacelike $t=const.$ hypersurfaces. 

 By substituting the spatial Ricci scalar corresponding to the metric (\ref{FR1}) into the action functional (\ref{1-1aaa}), we derive the Arnowitt--Deser--Misner (ADM) action \cite{Arnowitt:1962hi} for the model in the presence of a cosmological constant $\Lambda$ as
\begin{align}
S_\text{ADM}=\int L_\text{ADM} \mathrm{d}t,
\end{align}
where 
\begin{align}\label{LagADM1}
\begin{split}
L_\text{ADM}&=\frac{3\mathcal V_kNa^{3}}{8\pi G}\biggl[-\frac{\Lambda}{3}+\frac{1}{N^2}\biggl(-\frac{{\dot a}^2}{a^2}+\frac{kN^2}{a^2}\biggl)\biggl] +\frac{\mathcal V_kNa^{3}}{2}\Bigg(\frac{\dot T^2}{N^2}-1\Bigg)\rho\\
&=\frac{9N}{8\sqrt{G}}\left[-\frac{4}{9N^2}\dot x^2-\frac{\Lambda}{3}x^2+\left(\frac{\mathcal V_k}{3\pi\sqrt{G}}\right)^\frac{2}{3}kx^\frac{2}{3} \right]+\frac{3\pi\sqrt{G}Nx^2}{2}\Bigg(\frac{\dot T^2}{N^2}-1\Bigg)\rho,
\end{split}
\end{align}
is the ADM Lagrangian, and we defined the new scale factor 
\begin{equation}
    \label{newscale}
    a=\left( \frac{3\pi\sqrt{G}}{\mathcal V_k}\right)^\frac{1}{3}x^\frac{2}{3},
\end{equation}
in the second equality.

The conjugate momenta of $x$, and $T$ are, respectively
\begin{equation}
    \label{momenta}
    \Pi_x=\frac{\partial L_\text{ADM}}{\partial {\dot x}}= -\frac{\dot x}{\sqrt{G}N},~~~~~\Pi_T=\frac{\partial L_\text{ADM}}{\partial {\dot T}}=\frac{3\pi\sqrt{G}x^2\rho\dot T}{N}.
\end{equation}

The Euler--Lagrange equation corresponding to mass density gives us the constraint $N=|\dot T|$. The last term in (\ref{LagADM1}) is the ADM Lagrangian of the dust, consequently, the super-Hamiltonian of the dust is given by
\begin{equation}
    \label{dustH}
H_\text{dust}=\Pi_T\dot T-\frac{3\pi\sqrt{G}}{2}\Bigg(\frac{\dot T^2}{N^2}-1\Bigg)\rho=\text{sgn}(\dot T)N\Pi_T=N\mathcal H_\text{dust},
\end{equation}
where in the second equality we used the constraint $N=|\dot T|$.
In addition, the super-Hamiltonian of the gravitational part is given by
\begin{align}
\begin{split}
H_\text{gravity}&=\dot x\Pi_x-\frac{9N}{8\sqrt{G}}\left[-\frac{\Lambda}{3}x^2+k\left(\frac{\mathcal V_k}{3\pi\sqrt{G}}\right)^\frac{1}{3}x^\frac{2}{3}-\frac{4}{9N^2}\dot x^2 \right]\\&=-N\left\{\frac{\sqrt{G}}{2}\Pi_x^2+\frac{9}{8\sqrt{G}}\left(\left(\frac{\mathcal V_k}{3\pi\sqrt{G}}\right)^\frac{1}{3}kx^\frac{2}{3}-\frac{\Lambda}{3}x^2\right) \right\}=N\mathcal H_\text{gravity}.
\end{split}
\label{HG444}
\end{align}
Thus, the ADM Hamiltonian of the model is \cite{Maeda:2015fna}
\begin{equation}
    \label{Hamil}
  \begin{split}  H_\text{ADM}=&H_\text{gravity}+H_\text{dust}=N(\mathcal H_\text{gravity}+\mathcal H_\text{dust})\\=&N\left\{-\frac{\sqrt{G}}{2}\Pi_x^2-\frac{9}{8\sqrt{G}}\left[k\left(\frac{\mathcal V_k}{3\pi\sqrt{G}}\right)^\frac{1}{3}x^\frac{2}{3}-\frac{\Lambda}{3}x^2\right]+\varepsilon\Pi_T \right\},
  \end{split}
\end{equation}
where $\varepsilon=\text{sgn}(\dot T)$.

Therefore, the corresponding phase space is the cotangent bundle
\begin{equation}
    \Omega=\{x, T, N, \Pi_x, \Pi_{T}, \Pi_N  \}.
\end{equation}
The ADM Lagrangian of the model (\ref{LagADM1}) (as all ADM Lagrangians in GR) is singular because the conjugate momenta $\Pi_N$ weakly vanishes $    \Pi_N=\frac{\partial L_\text{ADM}}{\partial\dot N}\approx0$,
meaning that the Legendre transformation
\begin{equation}
 \{x, T, N, \dot x, \dot T, \dot N  \}\rightarrow   \{x, T, N, \Pi_x, \Pi_T, \Pi_N  \},
\end{equation}
is a surjective map and not an invertible map. Thus, we are working with a constrained Hamiltonian system since the Legendre transform is non-invertible. According to Dirac constraint theory \cite{doi:10.1142/8540}, the fact that $\Pi_N$ is weakly vanishes means that $N$, the lapse function is a freely selectable and physically irrelevant variable. Thus, the primary Hamiltonian of the model is $H_{\text{Prim}}=H_\text{ADM}+\lambda\Pi_N$, where $\lambda$ is a Lagrange multiplier. Now, we have
\begin{equation}
    \dot\Pi_N=\{\Pi_N, H_{Pr}\}=\mathcal H,
    \end{equation}

\noindent
for a certain quantity $\mathcal H$, which demonstrates that the primary constraint is not preserved, necessitating the imposition of the following secondary \cite{doi:10.1142/8540} super-Hamiltonian constraint
\begin{equation}
    \label{HRRR}
    \mathcal H=-\frac{\sqrt{G}}{2}\Pi_x^2-\frac{9}{8\sqrt{G}}\left[k\left(\frac{\mathcal V_k}{3\pi\sqrt{G}}\right)^\frac{1}{3}x^\frac{2}{3}-\frac{\Lambda}{3}x^2\right]+\varepsilon\Pi_T\approx0.
\end{equation}

The equations of motion are given by
\begin{equation}\label{Eq1}
    \begin{split}
        \dot x=&\frac{\partial H_\text{ADM}}{\partial \Pi_x}=-N\sqrt{G}\Pi_x,~~~~~\dot \Pi_x=-\frac{\partial H_\text{ADM}}{\partial x}=-\frac{3N\Lambda x}{4\sqrt{G}}+\frac{3k\left(\frac{\mathcal V_k}{3\pi\sqrt{G}}\right)^\frac{1}{3}}{4\sqrt{G}x^\frac{1}{3}},\\
        \dot T=&\frac{\partial H_\text{ADM}}{\partial \Pi_T}=N\varepsilon,\hspace{2.2cm}\dot \Pi_T=-\frac{\partial H_\text{ADM}}{\partial T}=0.
    \end{split}
\end{equation}

 We are interested in the spatial case of flat, $k=0$, AdS space. For this situation, the solutions for pair of $(T,\Pi_T)$, for choosing $\varepsilon=-1$, are
\begin{equation}\label{sol1}
    T=-t,~~~~\Pi_T=\text{const.}
\end{equation}
Thus, we may use the scalar field $T$ as a cosmic clock. Regarding the definition of $\Pi_T$ in (\ref{momenta}), the value of $\Pi_T$ is equal to the negative of the total mass of the dust in the comoving total volume $M=3\pi\sqrt{G}x^2\rho=\mathcal V_ka^3\rho$
\begin{equation}\label{sol2}
    \Pi_T=3\pi\sqrt{G}x^2\dot T\rho=-M.
\end{equation}
The dust effectively provides a cosmic time, therefore, there is a direct relation
between unitarity and geodesic completeness of the resulting spacetime in this time
parameter; if quantum dynamics do not remove classical singularities, loss of
unitarity occurs precisely at the singular points. In addition, the solution of (\ref{Eq1}) for the scale factor $x$ is
\begin{equation}
    \label{sol3}
    x(t)=\sqrt{\frac{8\sqrt{G}M}{3|\Lambda|}}\sin\left(\frac{\sqrt{3|\Lambda|}}{2}t\right).
\end{equation}
Note that the amplitude of the solution is fixed by the super-Hamiltonian constraint (\ref{HRRR}).
%%%%%%%%%%%%%%%%%%%%%%%%%%%%
\section{The Wheeler--DeWitt equation}\label{WDW}

Utilizing the minisuperspace quantization map in the configuration space $x\to {\hat{x}} = x$, $\Pi_x\to {\hat \Pi}_x=-i \partial/\partial x$ and $\Pi_T\to {\hat \Pi}_T=-i\partial/\partial T$ the super-Hamiltonian constraint (\ref{HRRR}) transforms to the WDW equation \cite{Jalalzadeh:2022bgz, Jalalzadeh:2022dlj}, where the super-Hamiltonian gives us the time evolution of the model
\begin{align}
i \frac{\partial \Psi}{\partial T}=\left\{-\frac{1}{2M_\text{P}}\frac{\partial^2}{\partial x^2}+\frac{9M_\text{P}}{8}\left[k\left(\frac{\mathcal V_kM_\text{P}}{3\pi}\right)^\frac{1}{3}x^\frac{2}{3}-\frac{\Lambda}{3}x^2\right]\right\}\Psi, \label{WdW444}
\end{align}
where $\Psi=\Psi(x,T)$ is the wavefunction, and $M_\text{P}=1/\sqrt{G}$ is the Planck mass in the natural units, $c=\hbar=k_B=1$.
This is the form of the Schr\"odinger equation where the scalar field $T$ acts as a time variable.

%In terms of $x$, the inner product is simply 
%\begin{equation}
%\langle \Phi|\Psi\rangle=\int^\infty_0\Phi^*\Psi \mathrm{d}x. \label{naiseki444}
%\end{equation}  

The WDW equation (\ref{WdW444}) is separable. Regarding Eq. (\ref{sol1}), the eigenvalue of $\hat\Pi_T$ is the total mass of the dust, i.e., $\hat\Pi_T\Psi=-M\Psi$. This leads us
\begin{equation}
    \label{WDW1122}
    \Psi(x,T)=e^{-iMT}\psi(x),
\end{equation}
where $\psi(x)$ satisfies the time-independent Schr\"odinger equation
\begin{eqnarray}\label{2-4}
\-\frac{1}{2M_\text{P}}\frac{\mathrm{d}^2\psi(x)}{\mathrm{d} x^2}+\frac{9M_\text{P}}{8}\left[k\left(\frac{\mathcal V_kM_\text{P}}{3\pi}\right)^\frac{1}{3}x^\frac{2}{3}-\frac{\Lambda}{3}x^2\right]\psi(x)=M\psi(x).
\end{eqnarray}

Note that the domain of the scale factor $a$ (and consequently $x$) is $[0,\infty)$, and our physical interpretation of quantum mechanical operators depend on their Hermicity and self-adjointedness, which is hampered by singular potentials or restricted domains of the wavefunction. However, in some cases, boundary conditions must be specified for recovering Hamiltonians to self-adjoint \cite{doi:10.1142/8540}. Thus, a satisfactory treatment of the WDW equation of our cosmological model requires solving the equation in a Hilbert space, and the solutions have to be associated with a self-adjoint operator. 
Thus there is a one-parameter, say $\gamma\in\mathbb R$,
1-parameter family of self-adjoint extensions of $\hat{\mathcal H}_\text{gravity}$ in (\ref{2-4}) on the half-line $x\geq0$, so the Hilbert space is the subspace wavefunctions that satisfy the Robin boundary condition
\begin{equation}\label{2-6}
\left(\frac{\mathrm{d}\psi(x)}{\mathrm{d}x}+\gamma\psi(x)\right)\Bigg|_{x\rightarrow0^+}=0,~~~~\gamma\in\mathbb R.
\end{equation}

The general square-integrable solution of equation (\ref{2-4}) with boundary condition
(\ref{2-6}) for a flat $(k=0)$ AdS spacetime is given by
\begin{multline}\label{2-7}
\psi(x)=
\frac{\sqrt{\pi}e^{-\frac{1}{2}M_\text{P}\omega x^2}}{2^{\frac{1}{4}-\frac{M}{2\omega}}\Gamma(\frac{3}{4}-\frac{M}{2\omega})}\Bigg\{{_1F_1}\left(\frac{1}{4}-\frac{M}{2\omega};\frac{1}{2};\frac{M_\text{P}\omega}{2}x^2\right)\\
-\frac{2\sqrt{M_\text{P}\omega}x\Gamma(\frac{3}{4}-\frac{M}{2\omega})}{\Gamma(\frac{1}{4}-\frac{M}{2\omega})}{_1F_1}\left(\frac{3}{4}-\frac{M}{2\omega};\frac{3}{2};\frac{M_\text{P}\omega}{2}x^2\right)\Bigg\},
\end{multline}
where $\omega:=\sqrt{3|\Lambda|/4}$ defined as the frequency of the oscillator and $_1F_1(\alpha;\beta;x)$ denotes the confluent hypergeometric
function. Making use of the properties, $_1F_1(\alpha;\beta;0)=1$ and $\frac{d}{dx}~_1F_1(\alpha;\beta;x)=\frac{\alpha}{\beta}\,\,_1F_1(\alpha+1;\beta+1;x)$,
 we can rewrite the boundary condition (\ref{2-6}) as
 \begin{equation}\label{2-8}
 \frac{\Gamma(\frac{3}{4}-\frac{M}{2\omega})}{\Gamma(\frac{1}{4}-\frac{M}{2\omega})}=\frac{\gamma}{2\sqrt{M_\text{P}\omega}}.
 \end{equation}
 
An unusual challenge arises when considering these extensions, as each introduces a distinct physics paradigm, indicating that the issue is not solely technical. When the parameter $\gamma$ is recognized to possess dimensions corresponding to the inverse of length, it emerges as a novel fundamental constant within the theoretical framework, as highlighted in references \cite{1986PhRT, Jalalzadeh:2024ncf, Moniz:2020emn}. However, as elucidated in \cite{Jalalzadeh:2017jdo}, the genesis of this undesired constant lies within the effective matter field Lagrangian featured in equation (28). If one were to substitute a ``real'' matter field, such as a scalar field or Maxwell's field Lagrangian, for $\rho$ in equation (28), the value of $\gamma$ would be constrained to only two permissible values: $\gamma=0$ (Dirichlet boundary condition) or $1/\gamma=0$ (DeWitt boundary or Neumann boundary condition).

{Now, let us show that the mass eigenvalues are the harmonic oscillator's energy spectrum for odd and even quantum numbers. To show this, not that in boundary condition (\ref{2-8}) for $\gamma=0$ we obtain $\Gamma(\frac{1}{4}-\frac{M}{2\omega})=0$. Also, for $1/\gamma=0$, we have $\Gamma(\frac{3}{4}-\frac{M}{2\omega})=0$. One can summarize these conditions as 
\begin{equation}
    \begin{cases}
        \frac{1}{4}-\frac{M}{2\omega}=-n_1,~~~~\text{if},~\gamma=0,\\
        \frac{3}{4}-\frac{M}{2\omega}=-n_1,~~~~\text{if},~\gamma=\pm\infty,
    \end{cases}
\end{equation}
where $n_1=0,1,2,...$. Thus,
\begin{equation}
    \begin{cases}
        M=\omega(2n_1+\frac{1}{2}),~~~~~~~~\text{if},~\gamma=0,\\
        M=\omega(2n_1+1+\frac{1}{2}),~~~\text{if},~\gamma=\pm\infty.
    \end{cases}
\end{equation}
Therefore, the total mass spectrum of the dust is }
\begin{equation}
    \label{mass}
    M=\omega\Big(n+\frac{1}{2}\Big)=\frac{3}{2}\sqrt{\frac{|\Lambda|}{3}}\Big(n+\frac{1}{2}\Big),
\end{equation}
{where $n=2n_1=0,2,4,...$, for $\gamma=0$, and $n=2n_1+1$ is an odd natural number for $1/\gamma=0$. Also, note that for these values of $M$, the confluent hypergeometric function reduces to the Hermite polynomials, which are eigenfunctions of the harmonic oscillator.} However, considering the solution for the semiclassical limit of equation (\ref{2-4}), note that the restriction for $\gamma$ of the possible values $\gamma=0$ or $1/\gamma=0$, due to the specified boundary conditions, becomes superfluous if we guarantee gamma with a finite value and $\omega$ large enough such that the right-hand side of (\ref{mass}) vanishes. 

If we assume the mass of the dust particle is $m$, with a total number of the particles in the comoving volume $\mathcal N$, then $M=\mathcal Nm$. This assumption, together with equation (\ref{mass}) leads us to rewrite the above equation in terms of the total number of particles.
We now estimate the total entropy of the dust-dominated universe. Let us assume that all the particles are identical. Then, the entropy of dust is equal to the total number of particles $S_\text{(dust)}=\mathcal N=M/m$ \cite{Rashki:2014noa}. Thus,
\begin{equation}\label{entropy1}
    S_\text{(dust)}=\frac{3}{2m}\sqrt{\frac{|\Lambda|}{3}}\Big(n+\frac{1}{2}\Big).
\end{equation}
Also, if we define the unit of entropy as
\begin{equation}
    \label{En11}
    \Delta S=S_\text{(dust)}(n+1)-S_\text{(dust)}(n)=\frac{3M_\Lambda}{2m},
\end{equation}
where $M_\Lambda=\sqrt{|\Lambda|/3}$, then one can rewrite (\ref{entropy1}) as
\begin{equation}
    \label{entropy2}
    S_\text{(dust)}=\Delta S\left(n+\frac{1}{2}\right).
\end{equation}

 Note that according to Refs. \cite{Rashki:2014noa, Ortiz:2005ak}, the entropy of dust in a closed universe filled with cosmic radiation and dust is $S_\text{(dust)}=\frac{\pi M_\text{P}}{m}\sqrt{24\big(n+\frac{1}{2}\big)}$. 
Therefore, while in a closed universe with dust and radiation fields, the total mass (or equivalently the entropy) is quantized in units of the Planck mass, in AdS spacetime, the total mass is quantized in units of the AdS energy scale.

\section{The fractional Wheeler--DeWitt equation}\label{FWDW}

As a non-local generalization of the WDW equation studied in the previous section, we assume and implement the following fractional quantization map \cite{Moniz:2020emn, Herrmann:2006nc, Rasouli:2021lgy, Jalalzadeh:2021gtq, Rasouli:2022bug}
\begin{equation}
    \label{map}
    (x,\Pi_x)\to \left(\hbar_\beta^{\beta-1}x^\beta,-i\hbar_\beta^{1-\beta}D^\alpha_x \right),%~~~(T,\Pi_T)\to (\hbar_\beta^{\beta-1}T^\beta,i\hbar_\beta^{1-\beta}D_T^\beta),
\end{equation}
where $\hbar_\beta$ is taken as a generalized coefficient with the dimension of the inverse of length (in natural units) and $D^\beta_x$
denotes a fractional derivative corresponding to the minisuperspace coordinate $x$ \cite{pozrikidis2018fractional}. Also, when $\beta = 1$, this operator simplifies to the standard operator in ordinary quantum mechanics. As we saw in equation (\ref{mass}), the quantum gravity effects are controlled by AdS energy scale $M_\Lambda=\sqrt{\frac{|\Lambda|}{3}}$. Thus, one can assume the generalized coefficient $\hbar_\beta$ % and $\hbar_\beta$ are 
is
\begin{equation}\label{adsscale}
    \hbar_\beta=M_\Lambda=\sqrt{\frac{|\Lambda|}{3}}.
\end{equation}

Utilizing the above fractional quantization map, the super-Hamiltonian constraint (\ref{HRRR}) transforms to the fractional WDW equation \cite{Jalalzadeh:2022bgz, Jalalzadeh:2022dlj}, where the matter's super-Hamiltonian gives us the time evolution of the model
\begin{equation}
i \partial_T\Psi(T,x)=H_\text{gravity}^{(\beta)}\Psi(T,x),
\label{WdWF}
\end{equation}
where
\begin{equation}\label{WdWF2}
 H_\text{gravity}^{(\beta)}=
    -\frac{1}{2M_\text{P}}M_\Lambda^{2(1-\beta)}D^{(\beta)}_xD^{(\beta)}_x+\frac{9M_\text{P}}{8}\left[k\left(\frac{\mathcal V_kM_\text{P}}{3\pi}\right)^\frac{1}{3}M_\Lambda^\frac{2(\beta-1)}{3}x^\frac{2\beta}{3}+M_\Lambda^{2\beta} x^{2\beta}\right].
\end{equation}

The Hermiticity of $D^{(\beta)}_xD^{(\beta)}_x$ depends on the specific choice of fractional derivative. Although the Caputo and Riemann definitions of the fractional derivative $D^{(\beta)}_xD^{(\beta)}_x$ are not Hermitian, using the Feller and Riesz definitions of the fractional derivative ensures the Hermiticity of this operator \cite{Laskin:2002zz}. The Riesz fractional derivative in $1D$ space is given by \cite{pozrikidis2018fractional}
\begin{equation}
    \label{sh5}
   D^{(\beta)}_xD^{(\beta)}_x\psi(x)=\frac{\alpha 2^{\alpha-1}}{\sqrt{\pi}}\frac{\Gamma(\frac{1+\alpha}{2})}{\Gamma(\frac{2-\alpha}{2})}\displaystyle\int_0^\infty\frac{\psi(x-v)-2\psi(x)+\psi(x+v)}{v^{\alpha+1}}{\mathrm d}v,
\end{equation}
 where $\alpha=\frac{\beta}{2}$ is the Lévy’s fractional parameter. For $\alpha=2$, Eq. (\ref{sh5}) coincides with the standard second-order derivative of $\psi(x)$. By a straightforward rearrangement, one can rewrite (\ref{sh5}) in the following form
\begin{equation}
    \label{sh5b}
 D^{(\beta)}_xD^{(\beta)}_x\psi(x)=\frac{\mathrm d}{\mathrm dx}D_x^{(\alpha)}\psi(x),  
\end{equation}
where the fractional first derivative is defined as \cite{pozrikidis2018fractional}
\begin{equation}
    \label{sh5c}
    D_x^{(\alpha)}=\frac{2^{\alpha-1}}{\sqrt{\pi}}\frac{\Gamma(\frac{1+\alpha}{2})}{\Gamma(\frac{2-\alpha}{2})}\int_0^\infty\frac{\psi(x+\nu)-\psi(x-\nu)}{\nu^\alpha}\mathrm{d}\nu.
\end{equation}
Note that, as the representation of the Riesz derivative in (\ref{sh5}) demonstrates, the Riesz fractional derivative is a nonlocal operator except when $\alpha=2$. This means that the fractional Laplacian in (\ref{WdWF2}) is influenced not just by $\psi(y)$ in proximity to $x$, but by $\psi(y)$ for all values of $y$. This nonlocality implies that when addressing the fractional WDW equation (\ref{WdWF2}), the form of the wavefunction in a given region depends not just on the potential in that region but on the
potential all of the minisuperspace. Furthermore, the impact of quantum gravity effects on classical domains cannot be overlooked. 

The fractional gravitational super-Hamiltonian (\ref{WdWF2}) is symmetric ( which is required for $H^{(\alpha)}_\text{gravity}$ to be self-adjoint) is
\begin{equation}
    \label{sy1}
\int_0^\infty \psi_1^*(x)H^{(\alpha)}_\text{gravity}\psi_2(x)dx=\int_0^\infty \psi_2(x)H^{(\alpha)}_\text{gravity}\psi_1^*(x)dx.
\end{equation}
This condition is equivalent to
\begin{equation}
    \label{sy2}
\int_0^\infty \psi_1^*(x)\frac{\mathrm{d}}{\mathrm{d}x}D_x^{(\alpha)}\psi_2(x)dx=\int_0^\infty \psi_2(x)\frac{\mathrm{d}}{\mathrm{d}x}D_x^{(\alpha)}\psi_1^*(x)dx,
\end{equation}
where the asterisk denotes complex conjugation. Integrating by parts twice the above condition gives us
\begin{equation}
\label{sy3}
\Big(\psi_1^*(x)D_x^{(\alpha)}\psi_2(x)-\psi_2(x)D_x^{(\alpha)}\psi_1^*(x) \Big)\Bigg|_0^\infty=0.
\end{equation}
 If we assume both $\psi(x)$ and $D_x^{(\alpha)}\psi(x)$ are square-integrable, the above relation leads us to the fractional generalization of the Robin boundary condition (\ref{2-6})
 \begin{equation}\label{2-6bb}
\Big(D_x^{(\alpha)}\psi(x)+\gamma\psi(x)\Big)\Bigg|_{x\rightarrow0^+}=0,~~~~\gamma\in\mathbb R.
\end{equation}
%On the other hand, in fractional quantum mechanics, $D_T^{(\beta)}$ is defined as left Caputa fractional derivative of order $\beta$, $D_T^{(\beta)}=i^{\beta-1}\partial_T^\beta$ \cite{2004JMP9N, 2008JMAA5D, 2017CSF16L}, defined by \cite{Caputo1967529}
%\begin{equation}
%    \label{Caputa}
 %   \partial_t^\beta f(t)=\frac{1}{\Gamma(1-\beta)}\int_0^t\frac{f'(\tau)}{(t-\tau)^\beta}\mathrm{d}\tau,~~~~~0<\beta\leq 1,
%\end{equation}
%where $f'(\tau)=\frac{\mathrm{d}f}{\mathrm{d}\tau}$, and $\Gamma(1-\beta)$ is the Gamma function. It is important to note that Naber \cite{2004JMP9N} has clearly and convincingly explained the rationale for raising the power of the imaginary unit $i$ to the order of the time derivative. The Caputo fractional derivative satisfies the following relations
%\begin{equation}
%    \label{Caputa2}
%    \begin{split}
%    &  \partial_t^\beta\{f(t)+g(t)\}=\partial_t^\beta f(t)+\partial_t^\beta g(t),\\
%  &  \partial_t^\beta t^k=\frac{k\Gamma(k)}{\Gamma(k+1-\beta)}t^{k-\beta},~~~
%   \partial_t^\beta c=0,
%    \end{split}
%\end{equation}
%where $c$ is a constant. For $\beta=1$ the Caputa fractional derivative reduces to the standard first-order derivative.

Given that $H_\text{gravity}^{(\alpha)}$ is time-independent, it is possible to solve equation (\ref{WdWF2}) using the method of separation of variables, by making the assumption     $\Psi(x, T)=f(T)\psi(x)$, equation (\ref{WdWF2}) separates into the following two equations
    \begin{equation}
 i\partial_Tf(T)=-Mf(T),\label{EE11}
\end{equation}
and
\begin{multline}
\Bigg\{-\frac{1}{2M_\text{P}}M_\Lambda^{2-\alpha}D^{(\alpha)}_xD^{(\alpha)}_x+\\\frac{9M_\text{P}}{8}\left[k\left(\frac{\mathcal V_kM_\text{P}}{3\pi}\right)^\frac{1}{3}M_\Lambda^\frac{\alpha-2}{3}x^\frac{\alpha}{3}+M_\Lambda^{\alpha} x^{\alpha}\right]\Bigg\}\psi(x)=M\psi(x).\label{EE22}
\end{multline}
Equation (\ref{EE22}) is the space fractional extension of (\ref{2-4}) with mass eigenvalue $M$.

%Utilizing the Laplace transform method, one can easily find the solution to the time-fractional equation (\ref{EE11}) as \cite{2008JMAA5D}
%\begin{equation}\label{Time32}    f(t)=E_\beta\left({M}{M_\Lambda^{\beta-1}(-iT)^\beta} \right), \end{equation}
%where 
%\begin{equation}
%    E_\beta(z)=\sum_{n=0}^\infty\frac{z^n}{\Gamma(\beta n+1)},
%\end{equation}
%is the Mittag--Leffler function, and we assumed $f(0)=1$. As \(\beta\) approaches 1, the Mittag-Leffler function turns into the ordinary exponential function, and the solution (\ref{Time32}) becomes $\exp(-iMT)$ as in the solution (\ref{WDW1122}).

Currently, there is no overarching solution for equation (\ref{sh5}) that explicitly includes a dependence on $\alpha$ \cite{Laskin:2002zz}. Therefore, we can use the Bohr--Sommerfeld quantization condition in a semiclassical limit. By integrating the exponential form of the wavefunction, $\psi(x) = \exp(-iS(x))$, into the Riesz fractional derivative (\ref{sh5}) and performing a Taylor series expansion of $\psi(x\pm\nu)$, we can employ the semiclassical approximation
\begin{equation}
    \label{H8b}
  D^{(\alpha)}_xD^{(\alpha)}_x\psi(x)=\frac{\alpha 2^{\alpha-1}}{\sqrt{\pi}}\frac{\Gamma((1+\alpha)/2)}{\Gamma((2-\alpha)/2)}e^{-iS}\int_0^\infty\frac{\sin^2(\frac{\nu}{2}\frac{dS}{dx})}{v^{1+\alpha}}d\nu=\Bigg|\frac{\mathrm dS}{\mathrm dx}\Bigg|^\alpha\psi(x).
\end{equation}
Using this relation, and $\frac{\mathrm dS}{\mathrm dx}=\Pi_x$ in equation (\ref{EE22}), gives us
\begin{equation}
    \label{H9}
   \frac{M_\Lambda^{2-\alpha}}{2M_\text{P}}|\Pi_x|^\alpha+\frac{9M_\text{P}}{8}\Bigg[k\left(\frac{\mathcal V_k}{3\pi\sqrt{G}}\right)^\frac{1}{3}M_\Lambda^\frac{\alpha-2}{3} x^\frac{\alpha}{3}+M_\Lambda^\alpha x^\alpha\Bigg]=M,
\end{equation}
where $\Pi_x$ is the canonical momentum corresponding to $x$.
 The classical turning points of this equation, i.e., $\Pi_x=0$, for a flat model, i.e., $k=0$ is $x= \bigg(\frac{4M}{M_\Lambda^\alpha M_\text{P}}\bigg)^\frac{1}{\alpha}$. Therefore, the Bohr--Sommerfeld quantization rule reads
\begin{equation}
    \label{BS}
   2 \int_0^{(\frac{4M}{M_\Lambda^\alpha M_\text{P}})^\frac{1}{\alpha}}\Pi_x\mathrm{d}x=2\pi(n+\frac{1}{2}),~~~~\text{(note that we assumed}~ \hbar=1),
\end{equation}
where $n$ is a large integer number. This gives us the fractional generalization of (\ref{mass})
\begin{equation}
    \label{Fmass}
     M^{(\text{fractional})}=\frac{3}{4}\left(\frac{\pi\alpha}{2B(\frac{1}{\alpha},1+\frac{1}{\alpha})}\right)^\frac{\alpha}{2}\sqrt{\frac{|\Lambda|}{3}}\left(n+\frac{1}{2}\right)^\frac{\alpha}{2},
\end{equation}
where $B(a,b)$ is a beta function. The fractional mass spectrum of the dust can be expressed in terms of the ordinary mass and the AdS energy scale, with equation (\ref{mass}) and equation (\ref{adsscale}), respectively
\begin{equation}
    \label{Fmass2}
     M^{(\text{fractional})}=\frac{3}{4}\left(\frac{\pi\alpha}{B(\frac{1}{\alpha},1+\frac{1}{\alpha})}\right)^\frac{\alpha}{2}M_{\Lambda}\Bigg(\frac{M}{M_{\Lambda}}\Bigg)^\frac{\alpha}{2}.
\end{equation}

The corresponding entropy is immediately obtained, once again assuming that the entropy of the fractional dust is equal to the number of particles with mass $m$, then $S^\text{fractional}=M^\text{fractional}/m$. We obtain for the entropy
or, in terms of the ordinary entropy (\ref{entropy1}) of the dust
\begin{equation}
    \label{Fentropy2}
     S^{(\text{fractional})}_\text{(dust)}=\frac{1}{2}\left(\frac{\pi\alpha}{2B(\frac{1}{\alpha},1+\frac{1}{\alpha})}\right)^\frac{\alpha}{2}\Delta S\left(\frac{S_\text{(dust)}}{\Delta S}\right)^\frac{\alpha}{2}.
\end{equation}
It is clear that by setting $\alpha$ to 2 in Eqs. (\ref{Fmass2}) and (\ref{Fentropy2}), we can readily obtain the standard mass (\ref{mass}) and entropy (\ref{entropy1}) of the dust.

Also, regarding the definition of the comoving total volume of the ordinary dust $M=3\pi\sqrt{G}x^2\rho=\mathcal V_ka^3\rho$, one can rewrite Eq. (\ref{Fmass2}) as
\begin{equation}
    \label{Fmass4}
    M^{(\text{fractional})}=\mathcal{V}_0^\text{(fractional)}\rho_\text{(fractional)}a^\frac{3\alpha}{2},
\end{equation}
where the fractional volume of flat space, $\mathcal{V}_0^\text{(fractional)}$, and the fractional dust density, $\rho_\text{(fractional)}$, are defined as
\begin{equation}
    \label{vol1}
    \mathcal{V}_0^\text{(fractional)}=\frac{3}{4}\left(\frac{\pi\alpha}{B(\frac{1}{\alpha},1+\frac{1}{\alpha})}\right)^\frac{\alpha}{2}\mathcal{V}_0^\frac{\alpha}{2},
\end{equation}
and
\begin{equation}
    \label{rhof}
    \rho_\text{(fractional)}=M_\Lambda^{1-\frac{\alpha}{2}}\rho^\frac{\alpha}{2}.
\end{equation}
Eq. (\ref{Fmass4}) demonstrates that the fractional dimension of the emergence classical universe in FQG is
\begin{equation}
    \label{dimension}
    D=\frac{3\alpha}{2},~~~~~~~\frac{3}{2}<D\leq3.
\end{equation}

Note that the $\alpha$ parameter is the quantity that provides the fractional scale for the mass and entropy spectrum. In the case of the mass spectrum, several models associate a fractal dimension with the dimension of the non-linear distributions of matter on the cosmic scale and in the description of the dynamics associated with galactic structures, which agrees with cosmological simulations and observations of the distribution of galaxies \cite{Gaite:2018kba, Gaite:2018lzc}.  Particularly, the relation $d_\text{mass}=3\alpha/2$ is in concordance with the reference \cite{Moradpour:2024uqa}, where the dynamical process of a collapsing and forming star is investigated from a fractional molecular cloud setup. Also, one can directly analyze the dimension of the mass distribution of the dust in terms of $\alpha$. When $\alpha=2$ ($D=3$) and, for example, in equation (\ref{Fmass4}), the volumetric scale of the distribution occurs, and the ordinary dimension case is recovered.

Finally, using Hamilton's equation for the fractional gravitational super-Hamiltonian, $\dot{x}={\partial}H_{\text{gravity}}^{(\alpha)}/{\partial}\Pi_x$, we can obtain the Friedmann equation of the minisuperspace associated model. In the comoving gauge $N=1$, for a flat Universe $k=0$, the super-Hamiltonian constraint is, from equation (\ref{H9})

\begin{equation}
\label{sHc}
  H_{\text{gravity}}^{(\alpha)}=-\frac{M_\Lambda^{2-\alpha}}{2M_\text{P}}|\Pi_x|^\alpha-\frac{9}{8}M_\text{P}M_\Lambda^\alpha x^\alpha+M=0.
\end{equation}

\noindent
The Friedmann equation is then obtained by Hamilton equations by

\begin{align}\label{ham1}
\dot{x}=&{\,}\frac{{\partial} H_{\text{gravity}}^{(\alpha)}}{\partial{\Pi_x}}=-\alpha\frac{M_\Lambda^{2-\alpha}}{2M_\text{P}}|\Pi_x|^{\alpha-1}\text{sgn}(\Pi_x);\\
\dot{\Pi_x}=&{\,}-\frac{{\partial} H_{\text{gravity}}^{(\alpha)}}{\partial{x}}=\frac{9}{8}{\alpha}M_\text{P}{M_\Lambda}^{\alpha}x^{\alpha-1},
\end{align}

\noindent
such that, solving equation (\ref{ham1}) for $|\Pi_x|^{\alpha}$ and substituting in equation (\ref{sHc}), we have

\begin{equation}
\label{Feq1}
  -\frac{1}{\alpha^{\frac{\alpha}{\alpha-1}}}\Bigg(\frac{M_\Lambda^{2-\alpha}}{2M_\text{P}}\Bigg)^{\frac{1}{1-\alpha}}|\dot{x}|^{\frac{\alpha}{\alpha-1}}-\frac{9}{8}M_\text{P}M_\Lambda^\alpha x^\alpha+M=0.
\end{equation}

\noindent
To visualize the result in conventional terms, we use the rescaled scale factor $a=a(t)$, 

\begin{equation}
\label{sfactor}
a=\Bigg(\frac{3\pi}{\mathcal{V}_{0}M_{\text{P}}}\Bigg)^{\frac{1}{3}}x^{\frac{2}{3}},
\end{equation}

\noindent
and in the regime $\dot{x}>0$, the Friedmann equation in terms of the Hubble parameter is

\begin{multline}
\label{Feq2}
  H^2= \Bigg(\frac{\alpha}{3}\frac{M_\Lambda^{2-\alpha}}{M_\text{P}}\Bigg)^2\Bigg[\Bigg(\frac{3\pi}{\mathcal{V}_{0}M_{\text{P}}}\Bigg)^{\frac{\alpha}{2(\alpha-1)}}Ma^{\frac{3\alpha}{2(1-\alpha)}}-\\\frac{9}{8}M_\text{P}M_{\Lambda}^{\alpha}\Bigg(\frac{\mathcal{V}_{0}M_{\text{P}}}{3\pi}\Bigg)^{\frac{\alpha(\alpha-2)}{2(\alpha-1)}}a^{\frac{3\alpha(\alpha-2)}{2(\alpha-1)}}\Bigg]^{\frac{2(\alpha-1)}{\alpha}}.
\end{multline}

 The $\alpha$-dependent modification of equation (\ref{Feq2}) introduces a coupling between the expansion rate $\dot{a}(t)$ and the scale factor that is less dependent on $a(t)$ than in the standard model of cosmology. Since $\dot{a}(t)$ changes faster and $a(t)$ changes slower with $\alpha<2$, the slower variation of $\dot{a}$ could imply that transitions between cosmological epochs (e.g., radiation to matter) might occur more gradually, potentially altering the structure formation rate. From equation (\ref{Feq2}), it should be noted that this modified Friedmann equation fully recovers its standard form with a negative $\Lambda$, when $\alpha=2$.

\section{Conclusion}\label{sec4man}

This paper investigated the consequences of a fractional quantization map in an AdS background FLRW universe using the Brown-Kucha\v{r} dust as the matter field, which acted as the time coordinate in minisuperspace. The classical and quantum dynamics of the matter field were obtained, and the WDW equation was demonstrated to resemble a Schrödinger-like equation for a quantum oscillator. The mass spectrum and entropy related to the dust were extracted using the semiclassical limit for a flat ($k=0$) AdS universe.

Here, the fractional quantization provided a generalized framework for examining the fractional WDW equation of the Brown--Kucha\v{r} dust, which led to a deeper understanding of the connection between its mass dimension $d_{\text{mass}}$ and the fractional parameter $\alpha$. The relation $d_\text{mass}=3\alpha/2$ was derived, establishing a direct link between Lévy’s fractional parameter and the fractal nature of the mass distribution. In agreement with reference \cite{Moradpour:2024uqa}, this result suggests that some systems, such as collapsing gas clouds or scalar fields in cosmic backgrounds, might have fractal qualities in their mass distribution, opening the door to a more expansive explanation of cosmological structures. The dynamics of these systems, both at the classical and quantum levels, may be accurately described by this fractal characteristic, which is defined by their fractional mass dimension.

\ack
S.J. acknowledges financial support from the National Council for Scientific
and Technological Development -- CNPq, Grant no. 308131/2022-3.
\vspace{.3cm}
\section*{References}
\bibliographystyle{iopart-num}       
\bibliography{FractionalADS}   % name your BibTeX data base

\providecommand{\newblock}{}
\begin{thebibliography}{10}
\expandafter\ifx\csname url\endcsname\relax
  \def\url#1{{\tt #1}}\fi
\expandafter\ifx\csname urlprefix\endcsname\relax\def\urlprefix{URL }\fi
\providecommand{\eprint}[2][]{\url{#2}}

\bibitem{Jalalzadeh:2020bqu}
Jalalzadeh S and Vargas~Moniz P 2022 {\em {Challenging Routes in Quantum Cosmology}\/} (World Scientific) ISBN 978-981-4415-06-4

\bibitem{Rovelli:2004tv}
Rovelli C 2004 {\em {Quantum gravity}\/} Cambridge Monographs on Mathematical Physics (Cambridge, UK: Univ. Pr.)

\bibitem{Oriti:2009zz}
Oriti D 2009 {\em {Approaches to quantum gravity: Toward a new understanding of space, time and matter}\/} (Cambridge University Press) ISBN 978-0-521-86045-1, 978-0-511-51240-7

\bibitem{Jalalzadeh:2022rxx}
Jalalzadeh S 2022 {\em Phys. Lett. B\/} {\bf 829} 137058 (\textit{Preprint} \eprint{2203.09968})

\bibitem{Penrose:1964wq}
Penrose R 1965 {\em Phys. Rev. Lett.\/} {\bf 14} 57--59

\bibitem{Hawking:1970zqf}
Hawking S~W and Penrose R 1970 {\em Proc. Roy. Soc. Lond. A\/} {\bf 314} 529--548

\bibitem{Hawking:1975vcx}
Hawking S~W 1975 {\em Commun. Math. Phys.\/} {\bf 43} 199--220 [Erratum: Commun.Math.Phys. 46, 206 (1976)]

\bibitem{Jalalzadeh:2023mzw}
Jalalzadeh S, Moradpour H and Moniz P~V 2023 {\em Phys. Dark Univ.\/} {\bf 42} 101320 (\textit{Preprint} \eprint{2308.12089})

\bibitem{Jalalzadeh:2022dlj}
Jalalzadeh S, Mohammadi A and Demir D 2023 {\em Phys. Dark Univ.\/} {\bf 40} 101227 (\textit{Preprint} \eprint{2210.02629})

\bibitem{Jalalzadeh:2022bgz}
Jalalzadeh S 2022 {\em Phys. Lett. B\/} {\bf 833} 137285 (\textit{Preprint} \eprint{2207.00727})

\bibitem{Trugenberger:2024miq}
Trugenberger C~A 2024 {\em Class. Quant. Gravity\/} (\textit{Preprint} \eprint{2409.09385})

\bibitem{Chen:2024iha}
Chen D~M and Wang L 2024 {\em Universe\/} {\bf 10} 333 (\textit{Preprint} \eprint{2409.02954})

\bibitem{Kruglov:2023rii}
Kruglov S~I 2023 {\em JHAP\/} {\bf 3} 53--58

\bibitem{Moniz:2020emn}
Moniz P~V and Jalalzadeh S 2020 {\em Mathematics\/} {\bf 8} 313 (\textit{Preprint} \eprint{2003.01070})

\bibitem{Brunetti:2022itx}
Brunetti R, Fredenhagen K and Rejzner K 2023 {\em {Locally Covariant Approach to Effective Quantum Gravity}\/} (Springer) pp 1--26 (\textit{Preprint} \eprint{2212.07800})

\bibitem{Kelly:2020uwj}
Kelly J~G, Santacruz R and Wilson-Ewing E 2020 {\em Phys. Rev. D\/} {\bf 102} 106024 (\textit{Preprint} \eprint{2006.09302})

\bibitem{Liu:2024soc}
Liu W, Wu D and Wang J 2024 {\em Phys. Lett. B\/} {\bf 858} 139052 (\textit{Preprint} \eprint{2408.05569})

\bibitem{Bojowald:2012xy}
Bojowald M 2012 {\em Class. Quant. Grav.\/} {\bf 29} 213001 (\textit{Preprint} \eprint{1209.3403})

\bibitem{Donoghue:2012zc}
Donoghue J~F 2012 {\em AIP Conf. Proc.\/} {\bf 1483} 73--94 (\textit{Preprint} \eprint{1209.3511})

\bibitem{Bertini:2024onw}
Bertini N~R, Rodrigues D~C and Shapiro I~L 2024 {\em Phys. Dark Univ.\/} {\bf 45} 101502 (\textit{Preprint} \eprint{2401.11559})

\bibitem{Battista:2023iyu}
Battista E 2024 {\em Phys. Rev. D\/} {\bf 109} 026004 (\textit{Preprint} \eprint{2312.00450})

\bibitem{Giacchini:2021abr}
Giacchini B~L, de~Paula~Netto T and Shapiro I~L 2022 {\em Nuovo Cim. C\/} {\bf 45} 34 (\textit{Preprint} \eprint{2111.14996})

\bibitem{LeonTorres:2023ehd}
Leon~Torres G, Garc\'\i{}a-Aspeitia M~A, Fernandez-Anaya G, Hern\'andez-Almada A, Maga\~na J and Gonz\'alez E 2023 {\em PoS\/} {\bf CORFU2022} 248 (\textit{Preprint} \eprint{2304.14465})

\bibitem{Shchigolev:2013jq}
Shchigolev V~K 2013 {\em Mod. Phys. Lett. A\/} {\bf 28} 1350056 (\textit{Preprint} \eprint{1301.7198})

\bibitem{Calcagni:2013yqa}
Calcagni G 2013 {\em JCAP\/} {\bf 12} 041 (\textit{Preprint} \eprint{1307.6382})

\bibitem{Calcagni:2017via}
Calcagni G 2017 {\em Phys. Rev. D\/} {\bf 96} 046001 (\textit{Preprint} \eprint{1705.01619})

\bibitem{Shchigolev:2021lbm}
Shchigolev V~K 2021 {\em Mod. Phys. Lett. A\/} {\bf 36} 2130014 (\textit{Preprint} \eprint{2104.12610})

\bibitem{Gonzalez:2023who}
Gonz\'alez E, Leon G and Fernandez-Anaya G 2023 {\em Fractal Fract.\/} {\bf 7} 368 (\textit{Preprint} \eprint{2303.16409})

\bibitem{Socorro:2023ztq}
Socorro J and Rosales J~J 2023 {\em Universe\/} {\bf 9} 185 (\textit{Preprint} \eprint{2302.07799})

\bibitem{Marroquin:2024ddg}
Marroqu\'\i{}n K, Leon G, Millano A~D, Michea C and Paliathanasis A 2024 {\em Fractal Fract.\/} {\bf 8} 253 (\textit{Preprint} \eprint{2402.13850})

\bibitem{Benetti:2024nxr}
Benetti F, Lapi A, Gandolfi G and Liberati S 2024 {\em Class. Quant. Grav.\/} {\bf 41} 175010 (\textit{Preprint} \eprint{2407.16787})

\bibitem{Trivedi:2024inb}
Trivedi O, Bidlan A and Moniz P 2024 {\em Phys. Lett. B\/} {\bf 858} 139074 (\textit{Preprint} \eprint{2407.16685})

\bibitem{Varao:2024eig}
Var\~ao G, Lobo I~P and Bezerra V~B 2024 {\em EPL\/} {\bf 148} 30001 (\textit{Preprint} \eprint{2405.13544})

\bibitem{Jamil:2011uj}
Jamil M, Momeni D and Rashid M~A 2012 {\em J. Phys. Conf. Ser.\/} {\bf 354} 012008 (\textit{Preprint} \eprint{1106.2974})

\bibitem{El-Nabulsi:2017vmp}
El-Nabulsi R~A 2017 {\em Int. J. Theor. Phys.\/} {\bf 56} 1159--1182

\bibitem{El-Nabulsi:2013mma}
El-Nabulsi R~A 2013 {\em Can. J. Phys.\/} {\bf 91} 618--622

\bibitem{Rasouli:2022bug}
Rasouli S~M~M, Costa E~W~O, Moniz P~V and Jalalzadeh S 2022 {\em {Fractal and Fractional}\/} {\bf 6} ISSN 2504-3110 (\textit{Preprint} \eprint{2210.00909}) \urlprefix\url{https://www.mdpi.com/2504-3110/6/11/655}

\bibitem{El-Nabulsi:2013mwa}
El-Nabulsi A~R 2013 {\em Indian J. Phys.\/} {\bf 87} 835--840

\bibitem{El-Nabulsi:2016dsj}
El-Nabulsi R~A 2016 {\em Rev. Mex. Fis.\/} {\bf 62} 240

\bibitem{El-Nabulsi:2017jss}
El-Nabulsi R~A 2017 {\em Can. J. Phys.\/} {\bf 95} 605--609

\bibitem{Rami:2015kha}
Rami E~N~A 2015 {\em Eur. Phys. J. Plus\/} {\bf 130} 102

\bibitem{El-Nabulsi:2017rdu}
El-Nabulsi R~A 2017 {\em Commun. Theor. Phys.\/} {\bf 68} 309

\bibitem{Landim:2021www}
Landim R~G 2021 {\em Phys. Rev. D\/} {\bf 103} 083511 (\textit{Preprint} \eprint{2101.05072})

\bibitem{Landim:2021ial}
Landim R~G 2021 {\em Phys. Rev. D\/} {\bf 104} 103508 (\textit{Preprint} \eprint{2106.15415})

\bibitem{Varieschi:2020ioh}
Varieschi G~U 2020 {\em Found. Phys.\/} {\bf 50} 1608--1644 [Erratum: Found.Phys. 51, 41 (2021)] (\textit{Preprint} \eprint{2003.05784})

\bibitem{Giusti:2020rul}
Giusti A 2020 {\em Phys. Rev. D\/} {\bf 101} 124029 (\textit{Preprint} \eprint{2002.07133})

\bibitem{Rasouli:2021lgy}
Rasouli S~M~M, Jalalzadeh S and Moniz P~V 2021 {\em Mod. Phys. Lett. A\/} {\bf 36} 2140005 (\textit{Preprint} \eprint{2101.03065})

\bibitem{Laskin:2002zz}
Laskin N 2002 {\em Phys. Rev. E\/} {\bf 66} 056108 (\textit{Preprint} \eprint{quant-ph/0206098})

\bibitem{metzler2000}
Metzler R and Klafter J 2000 {\em Physics Reports\/} {\bf 339}(1) 1--77

\bibitem{rocco1999}
Rocco A and West B~J 1999 {\em Physica A\/} {\bf 265}(3-4) 535--546

\bibitem{butera2014}
Butera S and Di~Paola M 2014 {\em Annals of Physics\/} {\bf 350} 146--158

\bibitem{tatom1995}
Tatom F~B 1995 {\em Fractals\/} {\bf 3} 217--229

\bibitem{Jalalzadeh:2021gtq}
Jalalzadeh S, da~Silva F~R and Moniz P~V 2021 {\em Eur. Phys. J. C\/} {\bf 81} 632 (\textit{Preprint} \eprint{2107.04789})

\bibitem{Jalalzadeh:2022uhl}
Jalalzadeh S, Costa E~W~O and Moniz P~V 2022 {\em Phys. Rev. D\/} {\bf 105} L121901 (\textit{Preprint} \eprint{2206.07818})

\bibitem{Junior:2023fwb}
Junior P~F~d~S, Costa E~W~d~O and Jalalzadeh S 2023 {\em Eur. Phys. J. Plus\/} {\bf 138} 862 (\textit{Preprint} \eprint{2309.12478})

\bibitem{deOliveiraCosta:2023srx}
de~Oliveira~Costa E~W, Jalalzadeh R, da~Silva Junior P~F, Rasouli S~M~M and Jalalzadeh S 2023 {\em Fractal Fract.\/} {\bf 7} 854 (\textit{Preprint} \eprint{2310.09464})

\bibitem{Jalalzadeh:2024qej}
Jalalzadeh R, Jalalzadeh S, Jahromi A~S and Moradpour H 2024 {\em Phys. Dark Univ.\/} {\bf 44} 101498 (\textit{Preprint} \eprint{2404.06986})

\bibitem{Moradpour:2024uqa}
Moradpour H, Jalalzadeh S and Javaherian M 2024 {\em Astrophys. Space Sci.\/} {\bf 369} 98 (\textit{Preprint} \eprint{2409.12869})

\bibitem{Brown:1994py}
Brown J~D and Kuchar K~V 1995 {\em Phys. Rev. D\/} {\bf 51} 5600--5629 (\textit{Preprint} \eprint{gr-qc/9409001})

\bibitem{Witten:1998zw}
Witten E 1998 {\em Adv. Theor. Math. Phys.\/} {\bf 2} 505--532 (\textit{Preprint} \eprint{hep-th/9803131})

\bibitem{Dong:2016eik}
Dong X, Harlow D and Wall A~C 2016 {\em Phys. Rev. Lett.\/} {\bf 117} 021601 (\textit{Preprint} \eprint{1601.05416})

\bibitem{DHoker:2002nbb}
D'Hoker E and Freedman D~Z 2002 {Supersymmetric gauge theories and the AdS / CFT correspondence} {\em {Theoretical Advanced Study Institute in Elementary Particle Physics (TASI 2001): Strings, Branes and EXTRA Dimensions}\/} pp 3--158 (\textit{Preprint} \eprint{hep-th/0201253})

\bibitem{Ryu:2006bv}
Ryu S and Takayanagi T 2006 {\em Phys. Rev. Lett.\/} {\bf 96} 181602 (\textit{Preprint} \eprint{hep-th/0603001})

\bibitem{Almheiri:2014cka}
Almheiri A and Polchinski J 2015 {\em JHEP\/} {\bf 11} 014 (\textit{Preprint} \eprint{1402.6334})

\bibitem{Faulkner:2013ica}
Faulkner T, Guica M, Hartman T, Myers R~C and Van~Raamsdonk M 2014 {\em JHEP\/} {\bf 03} 051 (\textit{Preprint} \eprint{1312.7856})

\bibitem{Harlow:2018tng}
Harlow D and Ooguri H 2021 {\em Commun. Math. Phys.\/} {\bf 383} 1669--1804 (\textit{Preprint} \eprint{1810.05338})

\bibitem{Strominger:1998yg}
Strominger A 1999 {\em JHEP\/} {\bf 01} 007 (\textit{Preprint} \eprint{hep-th/9809027})

\bibitem{Maldacena:1997re}
Maldacena J~M 1998 {\em Adv. Theor. Math. Phys.\/} {\bf 2} 231--252 (\textit{Preprint} \eprint{hep-th/9711200})

\bibitem{Basteiro:2022usu}
Basteiro P, Elfert J, Erdmenger J and Hinrichsen H 2022 {\em J. Phys. A\/} {\bf 55} 364002 (\textit{Preprint} \eprint{2201.10870})

\bibitem{Skvortsov:2017ldz}
Skvortsov E~D and Tran T 2017 {\em Universe\/} {\bf 3} 61 (\textit{Preprint} \eprint{1707.00758})

\bibitem{doi:10.1142/8540}
Jalalzadeh S and Vargas~Moniz P 2022 {\em {Challenging Routes in Quantum Cosmology}\/} (World Scientific) ISBN 978-981-4415-06-4

\bibitem{Arnowitt:1962hi}
Arnowitt R~L, Deser S and Misner C~W 2008 {\em Gen. Rel. Grav.\/} {\bf 40} 1997--2027 (\textit{Preprint} \eprint{gr-qc/0405109})

\bibitem{Maeda:2015fna}
Maeda H 2015 {\em Class. Quant. Grav.\/} {\bf 32} 235023 (\textit{Preprint} \eprint{1502.06954})

\bibitem{1986PhRT}
{Tipler} F~J 1986 {\em Phys. Rep.\/} {\bf 137} 231--275

\bibitem{Jalalzadeh:2024ncf}
Jalalzadeh S, Moradpour H and Tebyanian H 2024 {\em Class. Quant. Grav.\/} {\bf 41} 165006 (\textit{Preprint} \eprint{2405.12314})

\bibitem{Jalalzadeh:2017jdo}
Jalalzadeh S, Capistrano A~J~S and Moniz P~V 2017 {\em Phys. Dark Univ.\/} {\bf 18} 55--66 (\textit{Preprint} \eprint{1709.09923})

\bibitem{Rashki:2014noa}
Rashki M and Jalalzadeh S 2015 {\em Phys. Rev. D\/} {\bf 91} 023501 (\textit{Preprint} \eprint{1412.3950})

\bibitem{Ortiz:2005ak}
Ortiz C, Socorro J, Tkach V~I, Torres J and Rosales J 2005 {\em J. Phys. Conf. Ser.\/} {\bf 24} 167--172

\bibitem{Herrmann:2006nc}
Herrmann R 2007 {\em J. Phys. G\/} {\bf 34} 607--626 (\textit{Preprint} \eprint{nucl-th/0610091})

\bibitem{pozrikidis2018fractional}
Pozrikidis C 2018 {\em The Fractional Laplacian\/} (Chapman and Hall/CRC)

\bibitem{Gaite:2018kba}
Gaite J 2019 {\em Adv. Astron.\/} {\bf 2019} 6587138 (\textit{Preprint} \eprint{1810.02311})

\bibitem{Gaite:2018lzc}
Gaite J 2018 {\em JCAP\/} {\bf 07} 010 (\textit{Preprint} \eprint{1803.07419})

\end{thebibliography}

\end{document}